# PERCEPTUAL LEARNED IMAGE COMPRESSION VIA END-TO-END JND-BASED OPTIMIZATION

*Farhad Pakdaman, Sanaz Nami, and Moncef Gabbouj*

Faculty of Information Technology and Communication Sciences, Tampere University, Finland

## ABSTRACT

Emerging Learned image Compression (LC) achieves significant improvements in coding efficiency by end-to-end training of neural networks for compression. An important benefit of this approach over traditional codecs is that any optimization criteria can be directly applied to the encoder-decoder networks during training. Perceptual optimization of LC to comply with the Human Visual System (HVS) is among such criteria, which has not been fully explored yet. This paper addresses this gap by proposing a novel framework to integrate Just Noticeable Distortion (JND) principles into LC. Leveraging existing JND datasets, three perceptual optimization methods are proposed to integrate JND into the LC training process: (1) Pixel-Wise JND Loss (PWL) prioritizes pixel-by-pixel fidelity in reproducing JND characteristics, (2) Image-Wise JND Loss (IWL) emphasizes on overall imperceptible degradation levels, and (3) Feature-Wise JND Loss (FWL) aligns the reconstructed image features with perceptually significant features. Experimental evaluations demonstrate the effectiveness of JND integration, highlighting improvements in rate-distortion performance and visual quality, compared to baseline methods. The proposed methods add no extra complexity after training.

## 1. INTRODUCTION

The widespread use of smart devices has resulted in increasing capture, transmission, and storage of images. Moreover, the flourishing popularity of social media such as YouTube, Instagram, and TikTok, has made image and video integral to daily life for billions. To handle the ever-increasing storage and networking demands, efficient image/video compression techniques remain essential [1].

**Perceptual optimization** is among such efficient approaches in image/video compression, which aims to enhance compression efficiency by eliminating imperceptible visual redundancies in multimedia content, aligning with the characteristics of the Human Visual System (HVS) [2][3]. Among perceptual methods, Just Noticeable Distortion (JND) stands out as a widely used method, discerning the upper limit of distortions imperceptible to the HVS. These methods optimize compression by voluntarily degrading the quality to a certain threshold which is not perceivable by HVS, to save on bitrate [4]. To accomplish this, recent JND-based methods use the available JND datasets [5][6] and apply machine learning techniques to predict the JND level based on quality indicators such as QP (Quantization Parameter) [7] or QF (Quality Factor) [8]. JND datasets provides QP or QF values along with corresponding JND images, enabling the training of data-driven approaches. Integrating JND into traditional image compression standards such as JPEG [8], HEVC [2], and VVC (intra) [9] has significantly improved compression by integrating. This process often involves setting appropriate parameters such as QP or QF, according to HVS [10]. However, such optimization has not been fully studied for the emerging Learned image Compression (LC) methods.

**Learned image Compression** substitutes all or parts of compression pipeline with learnable structures, such as Deep Neural Networks (DNNs). This often involves an end-to-end training process to learn the best encoding and decoding parameters, enabling an adaptive and data-driven approach. Ballé et al. [11][12] proposed to use additive uniform noise to imitate quantization, to facilitate end-to-end training. Diverse network architectures, such as operational neural networks [13] or transformers [14] were explored to achieve high-quality image reconstruction. Advanced context models have been designed to capture latent spatial dependencies, and guide the entropy coding [10][15][16][17]. These advanced techniques improve compression at the cost of high computational complexity [18]. Typically, all these methods employ a loss function according to the rate-distortion optimization, where MS-SSIM or MSE is used to measure distortion. However, such a loss function predominantly aims at optimizing the signal fidelity of compressed images, neglecting crucial aspects of visual quality.

Although LC methods can be directly optimized for any given criteria, **perceptual optimization for LC** has not been fully investigated. Few early solutions highlighted the pivotal role of the loss function in determining overall performance. Consequently, they proposed perceptual loss functions to better align with HVS. Patel et al. [19] proposed a loss function that computes MS-SSIM between the original and reconstructed images, while employing MSE to measure the difference between the features extracted from the original and reconstructed images using a pretrained network. Chen et al. [20] developed a network that mimics perceptual image quality measurements, such as SSIM, PSNR, MS-SSIM, VIF, and VMAF, between original and reconstructed images, incorporating it into the loss function. Mohammadi et al. [21] investigated the perceptual impact of various image quality metrics applicable to the loss function. Following a subjective test, they concluded that DISTS and MS-SSIM exhibited the best overall performance. Ding et al. [22] defined a distortion-aware adjustor, using the ReLU activation function to determine distortion levels. For small distortions, the ReLU-based mechanism assigns zero distortion, as it is imperceptible to the HVS.

Most existing solutions optimize LC based on image fidelity and ignore important characteristics of HVS, and the few early perceptual solutions, do not fully investigate JND-based compression. Design of JND-based optimization for LC is currently an open problem, and challenging. Due to the changes between LC and traditional compression, existing solutions cannot be directly applied to LC. For instance, most LC methods do not yet provide a rate control mechanism such as QP, and hence, optimizing via estimating a QP is not an option here. Moreover, training LC methods require large datasets, and the limited size of JND datasets is a challenge.

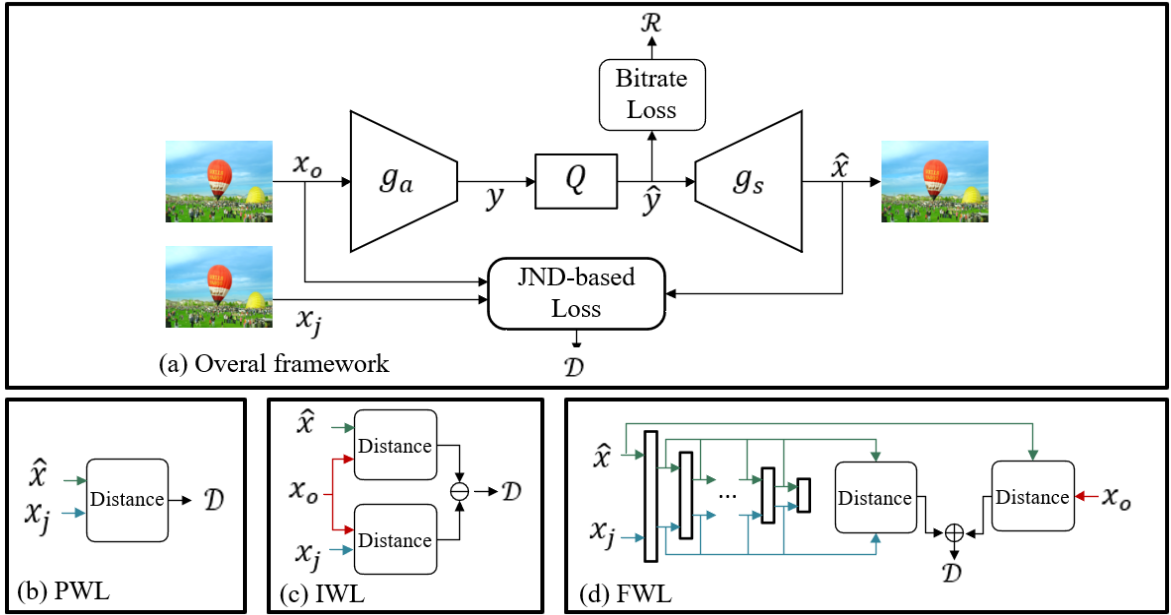

Fig. 1. The proposed learned image compression framework and JND-based perceptual loss functions. (a) overall framework. (b) Pixel-wise Loss (PWL). (c) Image-wise Loss (IWL). (d) Feature-wise Loss (FWL).

This paper tackles the above-mentioned challenges, by proposing JND-based optimization for LC. The proposed framework applies JND optimization at network training time, and integrates it into the Rate-Distortion loss. Using the existing JND datasets, network parameters are guided towards preserving HVS-related features, and discarding unperceived details. To this end, three JND-based perceptual loss functions have been proposed and evaluated, which are elaborated here.

(1) **Pixel-Wise JND Loss (PWL)**: This loss targets direct learning of the JND-quality image instead of the original (uncompressed) image. This approach promotes image fidelity between the JND-quality and decompressed images, intentionally disregarding the original image. The rationale is that the JND-quality images and their uncompressed images have similar visual quality, but they have less details and hence, should be easier to compress.

(2) **Image-Wise JND Loss (IWL)**: This loss quantifies the overall difference between the JND-quality image and its original counterpart. This difference is used to guide the optimization, to tolerate distortion levels below JND level.

(3) **Feature-Wise JND Loss (FWL)**: Utilizing a well-pretrained network, this loss calculates the dissimilarity between the features extracted from the JND-quality image and the decompressed image. Jointly emphasizing feature-level and pixel-level differences, this loss encourages the network to generate reconstructions that closely align with the perceptually significant characteristics of JND-quality image.

Moreover, a training strategy is proposed which jointly benefits from large (but unlabeled) raw image datasets, and limited JND-labeled datasets, to enable robust training based on JND. Accordingly, the following are the contributions of this paper:

- A JND-based optimization framework for LC is proposed, which achieves bitrate saving in reaching the JND quality.
- Three loss functions are designed to integrate JND into training of LC. These losses explore pixel-wise, image-wise, and feature-wise similarity for JND-based optimization.
- Extensive experiments evaluate different aspects of the proposed framework, and quantify the effectiveness.

The rest of the paper is organized as follows. Section 2 outlines the proposed methods, Section 3 presents and discusses the evaluation results, and Section 4 concludes the paper.

## 2. PROPOSED FRAMEWORK

Fig. 1(a) illustrates the overall framework, where a JND-based perceptual loss function is integrated into the LC training pipeline. First, the input original (uncompressed) image $x_o$ undergoes transformation through the parametric analysis encoder $g_a$, resulting in latent representation $y$. Then, the synthesis transform decoder $g_s$ uses the quantized latent representation $\hat{y}$ to reconstruct the decompressed image $\hat{x}$. This entire process can be represented as follows.

$$y = g_a(x_o; \phi_a) \quad (1)$$

$$\hat{y} = Q(y) \quad (2)$$

$$\hat{x} = g_s(\hat{y}; \phi_s) \quad (3)$$

Here, $\phi_a$, and $\phi_s$ are the optimized parameters of analysis encoder, and synthesis decoder, respectively. The encoder-decoder network is trained end-to-end, with a rate-distortion optimization which is formulated as (4):

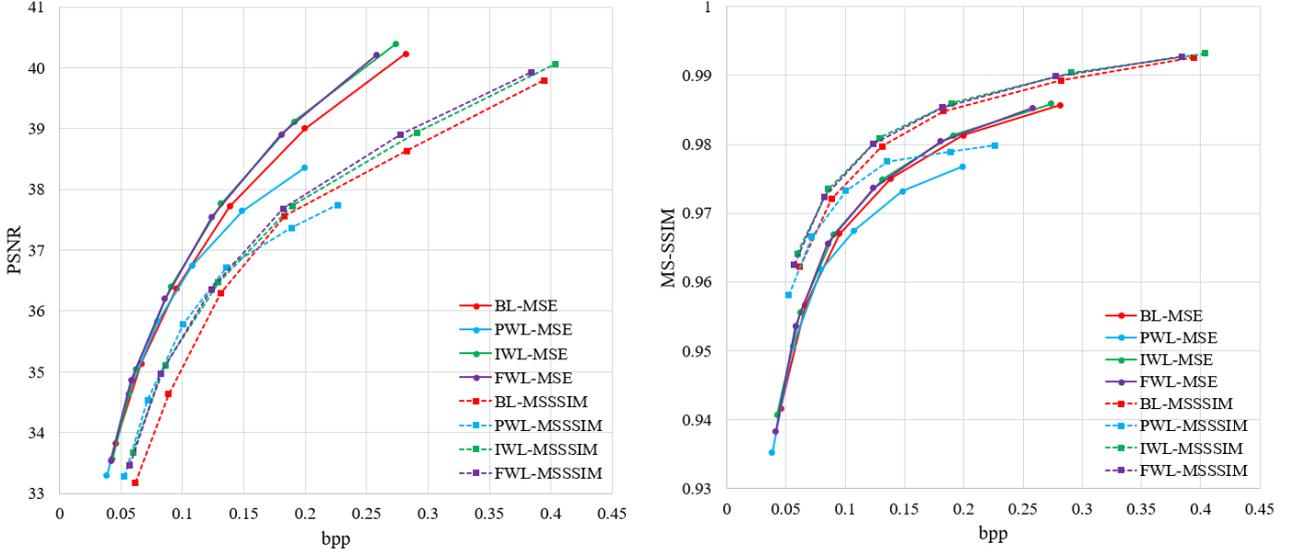

Fig. 2. Performance evaluation on VideoSet dataset

$$\mathcal{L} = \mathcal{R}(\widehat{\boldsymbol{y}}) + \lambda \cdot \mathcal{D} \qquad (4)$$

where, the Lagrange multiplier, $\lambda$, sets a balance between the rate $\mathcal{R}$ and distortion the $\mathcal{D}$. While the rate is approximated as the entropy of $\widehat{\boldsymbol{y}}$, most baseline LC methods define the distortion $\mathcal{D}$ via MSE or MS-SSIM between the original image $\boldsymbol{x_o}$ and the decompressed image $\widehat{\boldsymbol{x}}$, according to (5).

$$\mathcal{D} = d(\boldsymbol{x_o}, \widehat{\boldsymbol{x}}), \text{ where } d = MSE \text{ or } 1 - MSSSIM \qquad (5)$$

The baseline distortion targets image fidelity and does not consider HVS characteristics. The proposed framework introduces three alternative loss functions, to guide the optimization towards JND-quality. The subsequent subsections detail these loss functions.

### 2.1. Pixel-Wise JND Loss (PWL)

As noted earlier, the original and JND-quality images have similar visual qualities, but the JND-quality image has lower details which is favored by compression methods. Hence, we propose to directly train the LC methods to learn the JND-quality images, instead of uncompressed images (see Fig. 1 (b)). The distortion $\mathcal{D}$ to achieve this loss function, can be formulated as (6).

$$\mathcal{D} = d(\boldsymbol{x_j}, \widehat{\boldsymbol{x}}) \qquad (6)$$

This indicates that the loss is calculated between the reconstructed, and the JND-quality image, $\boldsymbol{x_j}$, instead of the original image.

### 2.2. Image-Wise Loss (IWL)

As previously emphasized, distortion levels lower or equal to the JND-quality images remain imperceptible to the HVS. IWL guides the model to tolerate this distortion level, by subtracting this level from the baseline distortion (see Fig. 1(c)). Distortion for IWL, can be formulated as (7), where the first term is the baseline loss, and the second term is the loss associated with the JND-quality image.

$$\mathcal{D} = d(\boldsymbol{x_o}, \widehat{\boldsymbol{x}}) - d(\boldsymbol{x_o}, \boldsymbol{x_j}) \qquad (7)$$

Unlike PWL with pixel-by pixel optimization, IWL applies the optimization on an image level by subtracting the overall JND loss.

### 2.3. Feature-Wise JND Loss (FWL)

An indirect way of guiding the network towards JND, is to guide the features to match those of a JND-quality image. Recent studies indicate that generic networks such as VGG, pretrained on large image datasets, learn features that perform well for various perceptual tasks [19]. Hence, this network is employed to compare reconstructed images with their JND-quality image, in a feature domain. Distortion for FWL is calculated by assessing the feature-level dissimilarity between the JND and reconstructed images. This loss is balanced by adding a pixel-level difference between the original and reconstructed images, as shown in Fig. 1 (d), and (8):

$$\mathcal{D} = \omega \cdot d(\boldsymbol{x_o}, \widehat{\boldsymbol{x}}) + (1 - \omega) \cdot d\left(F(\widehat{\boldsymbol{x}}), F(\boldsymbol{x_j})\right) \qquad (8)$$

where $d$ is calculated as MSE, $F$ represents the extracted features from VGG, and $\omega$ is a balancing weight.

### 2.4. Training Details

Training LC models require large image datasets. Existing JND datasets such as VideoSet [5] are not large enough for LC to generalize. Our experiments show that even training on large compression datasets such as JPEG-AI [23], and fine tuning on JND datasets tend to overfit the model. To remedy this issue, we mix same number of images randomly selected from JPEG-AI, with frames extracted from VideoSet. As JPEG-AI does not include JND labels, the original images are used as the JND-quality image for JPEG-AI. This balances between perceptual optimization and generalization.

### 3. EXPERIMENTAL RESULTS

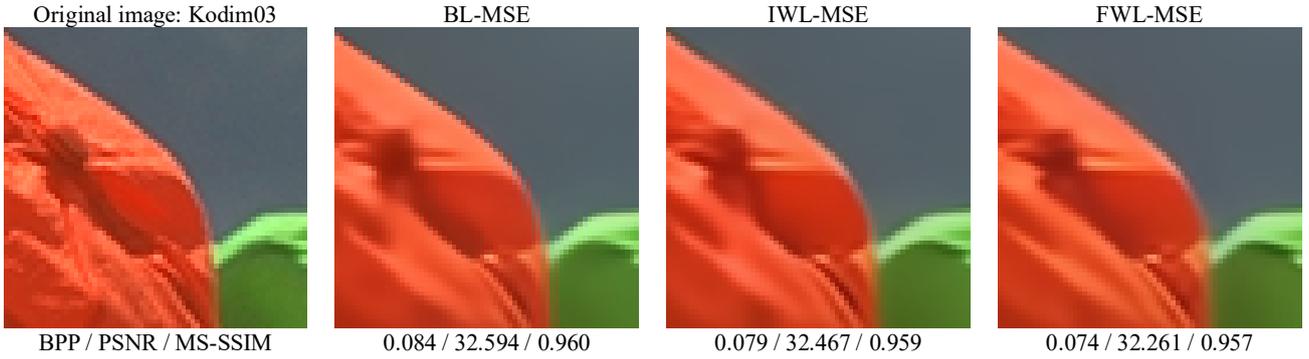

Fig. 3. Visual comparison of reconstructed images achieved by methods optimized with MSE

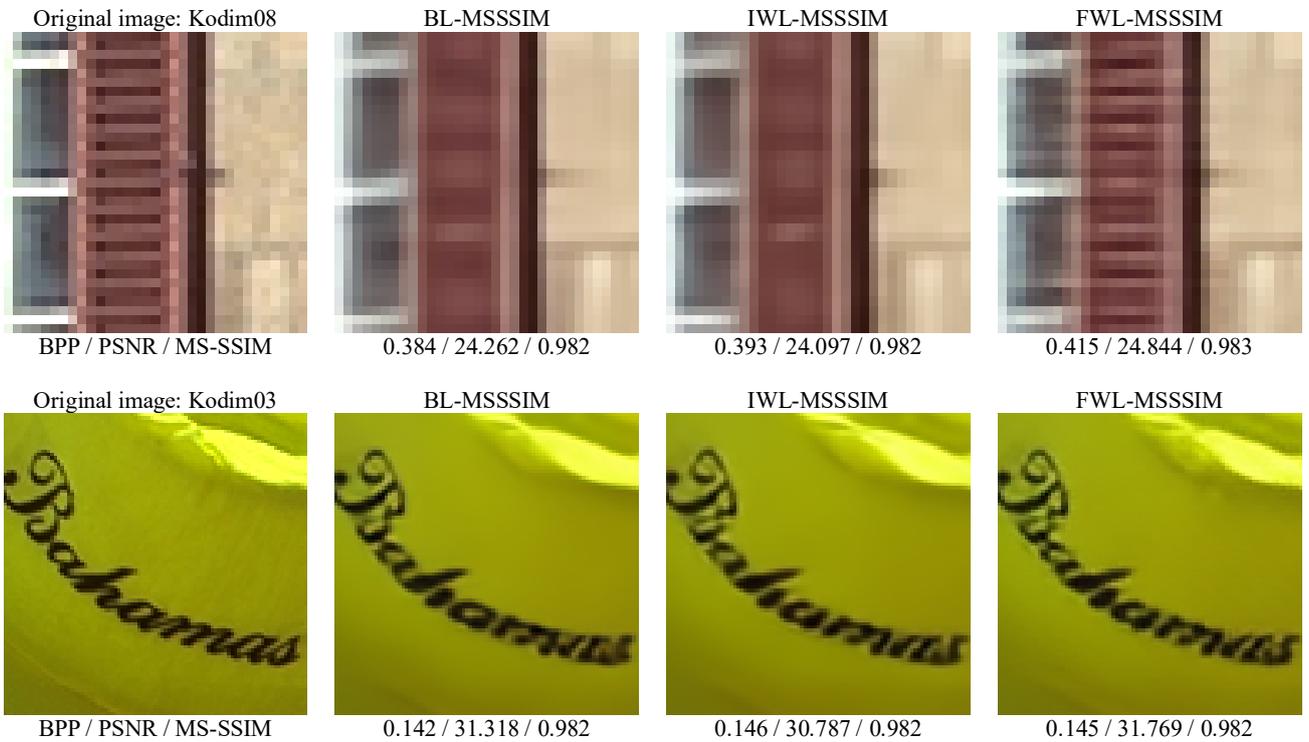

Fig. 4. Visual comparison of reconstructed images achieved by methods optimized with MS-SSIM

The proposed framework is implemented using the CompressAI [24] PyTorch library. The proposed methods are trained by initializing with the pre-trained parameters of Cheng-2020-attention method [10], which serves as the baseline. Our models are trained under six different quality levels. All models are trained once with MSE loss and once with MS-SSIM loss. For MSE-based models, $\lambda$ is set to 0.0018, 0.0035, 0.0067, 0.0130, 0.0250, and 0.0483, and for MS-SSIM-based models to 2.40, 4.58, 8.73, 16.64, 31.73, and 60.50. All models are trained for 100 epochs with a batch size of 16, and an initial learning rate of 1e-4. The proposed methods are trained using a mix of VideoSet [5] and JPEG-AI [23] as detailed in 2.4. Random patches of 256×256 are used during training. For the training phase, 80% of VideoSet is used, while 20% is reserved for evaluation. Moreover, we use the Kodak dataset [25] for qualitative analysis. Detailed experimental results are provided in next subsections.

### 3.1. Rate-Distortion Performance

To evaluate the Rate-Distortion (RD) performance, all test images were compressed and decompressed by each model. The rate is measured in terms of bits per pixel (bpp). Quality is measured with PSNR and MS-SSIM. Subsequently, RD curves are plotted on test set of VideoSet, depicted in Fig. 2 to compare the proposed methods against the baseline method.

For methods optimized for MSE, evaluating PSNR (Fig. 2 - left) shows competitive results at lower bitrates, close to the baseline method. At higher bitrates, two proposed methods, IWL-MSE and FWL-MSE outperform the baseline method. The same methods optimized for MS-SSIM consistently yields the best performance across all bitrates. However, it is noted that methods optimized for MS-SSIM gain lower PSNRs compared to MSE-based optimization, as expected. In terms of MS-SSIM, IWL-MSSSIM and FWL-MSSSIM, surpass the baseline again, emphasizing their effectiveness in preserving perceptual quality.

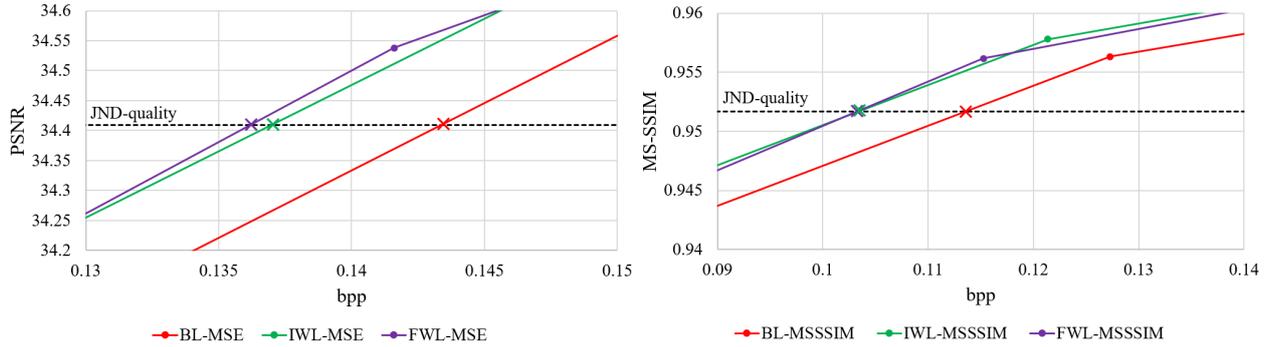

Fig. 5. Performance of methods based on JND-level quality for SRC005 example on VideoSet

Table I. Performance evaluation on VideoSet dataset based on BD-rate

| Method | IWL-MSE | | FWL-MSE | | IWL-MSSSIM | | FWL-MSSSIM | |
|---|---|---|---|---|---|---|---|---|
| BD-rate metric | PSNR | MS-SSIM | PSNR | MS-SSIM | PSNR | MS-SSIM | PSNR | MS-SSIM |
| VideoSet | -5.40 | -4.10 | -5.80 | -4.10 | -8.30 | -8.30 | -10.00 | -7.50 |

Table II. Average bitrate saving to achieve JND quality ($BS_{JND}$) for the test set of VideoSet

| metric | PSNR | | MS-SSIM | |
|---|---|---|---|---|
| method | IWL-MSE | FWL-MSE | IWL-MSSSIM | FWL-MSSSIM |
| $BR_{JND}$ (%) | -9.22 | -9.25 | -13.94 | -10.77 |

The high performance of IWL and FWL is justified by the fact that both methods relax the constraint of the distortion, hence, the training effort can be steered to further reduce the bitrate (entropy) for a given quality level. The proposed PWL on the other hand, achieves the lowest performance in most cases. It is concluded that the exclusive focus on pixel-wise difference was not sufficient to guide the training, and a direct optimization to learn JND cannot be as successful. Based on this observation, only IWL and FWL are selected for further evaluations.

Moreover, the R-D performance of each method against the baseline is quantified using the Bjontegaard-Delta bitrate (BD-rate), and summarized in Table I. BD-rate measures the average reduction in bitrate given similar quality, with negative values indicating bitrate saving. Notably, the proposed methods, IWL and FWL, achieve bitrate reduction ranging from 4.10% to 10%.

### 3.2. Qualitative Results

A comparison of the visual quality achieved by proposed and baseline methods in similar bitrates is provided in this section through visualizations of decompressed images. Fig. 3, shows an example where the proposed IWL and FWL optimized with MSE, exhibit slightly lower PSNR and MS-SSIM values compared to the baseline method (also lower bpp). However, they effectively overcome the aliasing distortion around the edges of baseline image, and achieve better visual quality. Similar degradation of baseline has been observed in more instances, for low bitrates. Fig. 4 provides two examples for MS-SSIM-based optimized methods, which further demonstrates the higher visual quality of IWL and FWL. They significantly outperform the baseline, showcasing enhanced visual quality and capturing more intricate details. This visual assessment confirms the perceptual improvements achieved by our proposed methods for LC.

### 3.3. JND-based Bitrate Saving

While the results in 3.1 and 3.2 shows the effectiveness of the proposed methods, they do not measure the efficiency in reaching JND level quality, which is a main goal of this research. To measure this, we introduce a JND-based Bitrate Saving metric, $BS_{JND}$. This metric quantifies the bitrate saving in reaching the JND-quality (in terms of PSNR or MS-SSIM), compared to the baseline method. $BS_{JND}$ is formulated as (9), where $bpp_{JND}^{BL}$ and $bpp_{JND}^{Proposed}$ represent the bpp of baseline and proposed method at JND-level quality, respectively.

$$BS_{JND} = \frac{bpp_{JND}^{Proposed} - bpp_{JND}^{BL}}{bpp_{JND}^{BL}} \times 100 \qquad (9)$$

To measure this, each test image from VideoSet was compressed with the baseline and the proposed methods to achieve the R-D curves. The JND quality for each test image is pre-known, as VideoSet is labeled with JND levels. Then, the bitrate to reach this JND quality is measured on the R-D curve for the baseline and the proposed method, and used in (9).

Fig. 5 shows one such example for SRC005 on VideoSet dataset. It is observed that for this image, both proposed methods reach JND level in lower bpps, compared to the baseline. Table II summarizes the average bitrate saving to reach JND for the test set, compared to the baseline. It can be observed that this saving is up to 9.25% and 13.94%, when measuring quality based on PSNR and MS-SSIM, respectively.

## 4. CONCLUSION

This paper presented a novel framework for perceptual optimization of learned image compression, through integration of JND principles. Leveraging existing JND datasets, three different loss functions have been proposed, to guide the LC training towards JND quality, with a pixel-wise, image-wise, or feature-wise similarity. Extensive experiments show improved performance and bitrate saving. Our findings reveal that (1) incorporating both the JND-quality image and original images as inputs to the loss function is more effective than using only JND images. (2) Optimizing for both pixel and feature similarity further enhances compression performance, resulting in images with improved visual quality. These outcomes underscore the effectiveness of our approach in achieving enhanced perceptual quality in LC methods. Based on the observations, future efforts shall be dedicated to investigating visual quality (such as JND) in LC images, as they can showcase different distortion types. Moreover, a next challenging step to fully benefit from JND models will be to enable the use of JND in the rate control mechanism for LC.

**ACKNOWLEGMENT**
This project has received funding from the European Union's Horizon 2020 research and innovation programme under the Marie Skłodowska-Curie grant agreement No [101022466].